# Observation of full momentum bandgap in photonic time crystals

Bolun Huang[1†], Zebin Zhu[1†§], Genrong Yu[1], Zhen Gao[1]*

[1]State Key Laboratory of Optical Fiber and Cable Manufacturing Technology, Department of Electronic and Electrical Engineering, Guangdong Key Laboratory of Integrated Optoelectronics Intellisense, Southern University of Science and Technology, Shenzhen 518055, China

[†]These authors contributed equally to this work.
[§*]Corresponding author. Email: zhuzb@sustech.edu.cn (Z.Z.); gaoz@sustech.edu.cn (Z.G.)

**The hallmark feature of photonic time crystals (PTCs) is the momentum bandgap, yet opening such a gap is extremely challenging, as it demands strong and rapid temporal modulation of the material properties. Recent theoretical advances have shown that resonance effects can substantially expand the momentum bandgap, and even give rise to a full (infinite) momentum bandgap spanning the entire momentum space. Despite these predictions, a full momentum bandgap has yet to be observed experimentally. Here, we report the first experimental observation of full momentum bandgaps in a microwave PTC. By enhancing the resonant effect, we demonstrate that the momentum bandgap can be drastically widened in a dynamically modulated microwave surface-plasmon transmission-line metamaterial, leading to tighter spatiotemporal field confinement and greater robustness against temporal disorder. Remarkably, using a dynamically modulated microwave coupled resonator metamaterial characterized by coupled-resonator optical waveguide dispersion, we achieve a full momentum bandgap spanning the entire momentum space, thereby enabling arbitrary spatial localization and temporal amplification of microwave fields. Our findings establish a unified experimental framework for expanding momentum bandgaps—up to an infinite extent—with minimal requirements on modulation strength and speed, thus paving a viable route toward the first experimental realization of PTCs at optical frequencies.**

**Introduction**

Photonic time crystals (PTCs)[1-7], the temporal analog of spatial photonic crystals (PCs)[8-10], are artificial materials whose electromagnetic properties (e.g., refractive index) are modulated periodically in time while remaining uniform in space. In contrast to spatial PCs featuring frequency bandgaps ($\omega$-gap) and spatially decaying modes, PTCs host momentum bandgaps ($k$-gap)[4,11-18] and temporally amplifying modes, enabling intriguing non-resonant light amplification[19] and emission[20,21], exotic light-matter interactions[22], frequency conversion with reduced phase-matching requirements[23-25], and temporal localization effects[26,27]. However, it is notoriously difficult to open a noticeable $k$-gap, especially at optical frequencies[17,18,28,29], as it requires ultrafast, order-unity modulation depths of refractive index (e.g., $\delta_n/n \approx 1$, with $n$ the refractive index) that are too extreme and far beyond the reach of modern optical technologies[30,31]. Consequently, although PTCs have been experimentally demonstrated at microwave frequencies[32-37], their realization in the optical regime remains a primary yet highly demanding goal. To alleviate the stringent requirements on modulation depth and speed for opening a discernible $k$-gap at optical frequencies, several compelling theoretical strategies have been proposed to expand the $k$-gap—and even achieve a full $k$-gap—under modest modulation conditions. These include strong resonance effects[38-43], nonuniform waves[44], and bound states in the continuum[45]. Despite these advances, the experimental observation of expanded or full $k$-gaps has thus far remained elusive, owing to the formidable challenges of achieving strong resonance or nonuniform waves in PTCs.

In this work, we report the first experimental observation of expanded and full $k$-gaps in PTCs operating at microwave frequencies. Using a dynamically modulated microwave surface-plasmon transmission-line metamaterial governed by spoof surface plasmon polariton (SSPP) dispersion[46,47]—herein referred to as an SSPP-PTC—we elucidate the theoretical mechanism underlying resonance-induced expansion of $k$-gaps and experimentally demonstrate that these $k$-gaps broaden substantially with increasing modulation frequency, a clear manifestation of the resonant effect. We further show that larger $k$-gaps give rise to tighter spatiotemporal field confinement and enhanced robustness against temporal disorder. Strikingly, by employing a dynamically modulated, microwave coupled resonator metamaterial characterized by coupled-resonator optical waveguide (CROW) dispersion—herein referred to as CROW-PTC—we experimentally realize a full $k$-gap[48] spanning the entire momentum space, thus allowing for arbitrary spatial localization and temporal amplification of microwave fields. Our results provide experimental validation of the resonance-based strategy for expanding $k$-gaps and chart a promising path toward the first realization of PTCs at optical frequencies via resonance.

**Results**

**Theoretical model for PTCs with normal, expanded, and full $k$-gaps**

We start from the time-domain Maxwell equations to analyze the band structure and the associated $k$-gaps of PTCs with different polarization responses[49]. Maxwell's equations relate the electric field and the electric displacement field through $\nabla \times \nabla \times \boldsymbol{E} + \mu_0 \frac{\partial^2 \boldsymbol{D}}{\partial t^2} = 0$, where $\boldsymbol{D} = \varepsilon_0 \boldsymbol{E} + \boldsymbol{P}$. The relation between the polarization $\boldsymbol{P}$ and the electric field $\boldsymbol{E}$ depends on the specific properties of the medium. In a PTC, the material parameter is periodically modulated in time, leading to the generation of harmonics and the interaction between different harmonic orders. In this case, using the Floquet theorem, both the electric field and the polarization can be expanded as $\boldsymbol{E} = \sum_n E_n e^{-i(\omega + n\Omega)t}$ and $\boldsymbol{P} = \sum_n P_n e^{-i(\omega + n\Omega)t}$, where $\omega$ and $\Omega$ are Floquet frequency and modulation frequency, respectively. Under the regime of weak harmonic modulation approximation, it is sufficient to retain only the coupling between the $n = 0$ and $n = -1$ Floquet sidebands[38]. The eigen-equation can be listed by (see Methods for detailed derivation):

$$\begin{pmatrix} F(\omega) & G[(\Omega - \omega)^2] \\ G(\omega^2) & F(\Omega - \omega) \end{pmatrix} \begin{pmatrix} E_0 \\ E_{-1} \end{pmatrix} = k^2 \begin{pmatrix} E_0 \\ E_{-1} \end{pmatrix}. \tag{1}$$

The diagonal terms $F(\omega)$ and $F(\Omega - \omega)$ describe the static photonic dispersion band and its replica shifted by a frequency $\Omega$, respectively, and the non-diagonal terms $G[(\Omega - \omega)^2]$ and $G(\omega^2)$ represent their coupling. At $\omega = \Omega/2$, two static bands cross and their coupling opens a $k$-gap, within which the eigenfrequencies form a pair of complex-conjugate values. The coupling strength determines both the $k$-gap width and the imaginary part of the eigenfrequencies inside the bandgap.

Conventional PTCs are typically realized in non-dispersion media, as shown in the top panel of Fig. 1a, where the polarization from electrons and positive charges responds instantaneously to an applied electric field, yielding a simple relation $\boldsymbol{P} \propto \boldsymbol{E}$. In such systems, the PTCs are usually implemented by applying a weak Floquet harmonic modulation to the electron density $N(t) =$ $N_0(1 + m\cos(\Omega t))$, where $m$ is the modulation depth, leading to a periodic modulation of the susceptibility. Since the relative modulation depth of the permittivity is of the same order as that of the electron density $N$, the coupling between the two crossed Floquet bands (grey solid lines) is extremely weak, resulting in only a tiny $k$-gap (cyan solid lines), as shown in the middle panel of Fig. 1a. Within it, a mode with a positive imaginary part of the eigenfrequency (cyan dotted lines) leads to the amplification of electromagnetic waves in the PTC (bottom panel of Fig. 1a).

One way to expand the $k$-gap is to introduce a Lorentz resonance in the PTCs, i.e., $\varepsilon(\omega) = \varepsilon_\infty + \omega_p^2/(\omega_r^2 - \omega^2)$[38]. As shown in the top panel of Fig. 1b, in this system, electrons oscillate harmonically around positive charges with a resonance frequency $\omega_r$, leading to the polarization dynamics $\partial^2 \boldsymbol{P}/\partial t^2 + \omega_r^2 \boldsymbol{P} = \varepsilon_0 \omega_p^2 \boldsymbol{E}$. After applying a weak Floquet harmonic modulation to the resonance frequency, $\omega_r(t) = \omega_{r0}(1 + m\cos(\Omega t))$, the coupling coefficient can be derived as $G(\omega) \propto (\omega_{r0}^2 - \omega^2)^{-1}$ (see Methods). When the frequency $\omega = \Omega/2$ approaches the Lorentz resonance $\omega_{r0}$, the two static Floquet dispersion curves (grey solid lines) cross and the coupling coefficient $G(\omega)$ diverges at a large $k$, opening an expanded (semi-infinite) $k$-gap (orange solid lines), as depicted in the middle panel of Fig. 1b. within this expanded $k$-gap, the wider bandgap and larger imaginary part of eigenfrequency (orange dotted lines) yield stronger wave localization and higher gain (bottom panel of Fig. 1b). However, since the two static Floquet bands (grey solid lines) exhibit a substantial frequency mismatch near $k = 0$, the resulting $k$-gap exists only on the large-$k$ side of momentum space. Consequently, it is impossible to achieve a full (i.e., infinite) $k$-gap that spans the entire momentum space.

To further open a full $k$-gap, we consider a Drude–Lorentz medium $\varepsilon(\omega) = \varepsilon_\infty - \omega_{p1}^2/\omega^2 + \omega_{p2}^2/(\omega_r^2 - \omega^2)$, as shown in the top panel of Fig. 1c. In this material, electrons not only participate in localized Lorentz resonances but also in collective longitudinal charge oscillations driven by the background potential. Consequently, the dispersion relation is governed by two characteristic resonance frequencies: the Lorentz resonance $\omega_r$ at $k \to \infty$, and the longitudinal collective plasmon frequency $\omega_{col}$ at $k = 0$. If both $\omega_{col}$ and $\omega_r$ are subjected to weak Floquet modulation with the same amplitude and phase, the non-zero collective plasmon frequency $\omega_{col}$ reduces the frequency mismatch near $k = 0$, and its modulation further enhances the coupling between modes in the small-$k$ region. As a result, such a PTC based on the Drude–Lorentz medium can support a full $k$-gap that spans the entire momentum space (middle panel of Fig. 1c). Within this full $k$-gap, as shown in the bottom panel of Fig. 1c, the real part of the eigenfrequency (purple solid line) forms a perfectly flat band, indicating that the corresponding modes are frozen in space at arbitrary positions and exhibit extreme spatial localization—approaching the atomic scale in principle. Furthermore, the eigenfrequencies also acquire a larger imaginary part (purple dotted lines), leading to more rapid exponential temporal amplification.

**Observation of expanded $k$-gaps in SSPP-PTC.**

To experimentally observe the resonance-induced expansion of $k$-gaps, we designed and fabricated a microwave-frequency SSPP-PTC. Figures 2a-c present the top and bottom views of the sample along with its unit cell. The top layer features periodic metallic strips spaced by $w = 4$ mm, each connected via a metallic via to a varactor diode on the bottom layer, thereby introducing a tunable distributed capacitance. A DC bias of 14 V is applied to set the average varactor capacitance to $C_{\text{var}} = 12$ pF, and a sinusoidal modulation voltage is applied to the varactors through a DC-block bandpass filter (BF), yielding an effective time-varying distributed capacitance $C_{\text{eff}}(t) = C_0(1 + m\cos\Omega t)$, where $m$ is the modulation depth and $\Omega$ the modulation frequency. The static part of the effective distributed capacitance $C_0 = 19.8$ pF incorporates the $C_{\text{var}}$, the BF's influence, and the amplifier's output resistance (see Methods for experimental setup and equivalent circuit). Figure 2d displays the measured (colormap) and theoretical (white dotted line) static SSPP dispersion, which terminates at the Brillouin zone boundary with a resonance frequency $\omega_{\text{r0}} = 2/\sqrt{(L_s - 2S)C_0}$, where $L_s$ is the equivalent coupling inductance of a metallic strip and $S$ represents the contribution of mutual inductance between adjacent strips (see Methods for the correction to the dispersion when the theoretical model includes mutual-inductance).

With the capacitance modulation depth fixed at $\Delta C = 2.38$ pF, we gradually increased the modulation frequency $\Omega$, bring $\Omega/2$ closer to the resonance frequency $\omega_{\text{r0}} = 371$ MHz. As expected, the $k$-gap expanded monotonically with increasing modulation frequency (Figs. 2e-g), and the spatial localization of the modes intensified accordingly (Figs. 2h-i). Specifically, for $\Omega = 575$ MHz, as shown in Figs. 2e and 2h, the resonant enhancement of the effective permittivity modulation depth remains limited because $\Omega/2$ lies well below the resonance frequency $\omega_{\text{r0}}$, resulting in a narrow $k$-gap characterized by low modal gain and weak spatiotemporal field confinement. Owing to the inherent transmission-line losses, these weakly amplified modes are largely suppressed and difficult to discern. The top panel of Fig. 2h shows that at $t = 70$ ns, the spatial distribution of $|H_z|^2$ remains diffusive, consistent with the behavior expected for a narrow *k*-gap.

When the modulation frequency was raised to $\Omega = 675$ MHz, $\Omega/2$ drew nearer to the resonance frequency $\omega_{r0}$, significantly enhancing the system's parametric response. This enhancement effectively increased the relative modulation depth of the effective permittivity, producing an expanded *k*-gap (Fig. 2f). A wider *k*-gap corresponds to a larger imaginary part of the eigenfrequency, which manifests as stronger exponential temporal amplification. Moreover, because the *k*-gap encompasses a wider momentum interval, the spatial localization of the mode field is correspondingly reinforced. As shown in Fig. 2i, compared with the case in Fig. 2h, the wave packet exhibits distinct exponential amplification and a markedly more localized spatial profile at $t = 70$ ns.

Upon further increasing the modulation frequency to $\Omega = 725$ MHz, $\Omega/2$ moved into even closer proximity to the resonance frequency $\omega_{r0}$, thereby maximizing the resonant enhancement. As a result, the *k*-gap broadened to the Brillouin zone boundary, forming an approximately semi-infinite *k*-gap (Fig. 2g). The corresponding spatiotemporal field distribution, presented in Fig. 2j, reveals that the mode experiences the most dramatic temporal amplification; at $t = 70$ ns, the spatial field confinement is further intensified, with the gain mode observable solely near the excitation source. The SSPP-PTC thus gives rise to an expanded *k*-gap that extends from a finite wavevector to the Brillouin zone boundary ($k = \pi/a$). The underlying physical mechanism is straightforward: as the Floquet frequency $\Omega/2$ approaches the resonance frequency, the resonance effect induces a divergent enhancement of the relative modulation depth, such that even modest harmonic modulation suffices to open a substantially larger *k*-gap.

**Observation of full *k*-gaps in CROW-PTC**

Note that in the above SSPP-PTC governed by SSPP dispersion, it is impossible to achieve a full *k*-gap spanning the entire Brillouin zone because the SSPP dispersion relation originates at zero frequency at *k* = 0. To further expand the *k*-gap to an infinite extent, we fabricated a CROW-PTC characterized by CROW dispersion, which begins at a nonzero frequency at *k* = 0. The top and bottom views of the experimental sample, along with its unit cell structure, are presented in Figs.

3a-c. Compared with the SSPP-PTC shown in Figs. 2a-c, the CROW-PTC introduces an additional resonance frequency $\omega_{col}$ at $k = 0$ by incorporating an inductor $L_0 = 9.2$ nH in parallel with the varactor diode, as illustrated in the insets of Figs. 3b and 3c. Meanwhile, the spacing between adjacent metallic strips on the top layer is increased to $w = 12$ mm, thereby weakening the coupling between neighboring resonators. In the experiment, a DC bias of 10.8 V was applied to fix the average varactor capacitance at $C_{var} = 17$ pF, which, together with the filter's influence and the amplifier's output resistance, determines the effective capacitance $C_{eff}(t) = C_0(1 + m\cos\Omega t)$, with $C_0 = 27.8$ pF. The inductor $L_0$ was connected in series with a large capacitor of 200 pF to provide DC isolation. The modulation signal was applied to the varactors through a DC-blocking BF, imposing a sinusoidal voltage with angular frequency $\Omega$ across them to achieve periodic temporal modulation.

Figure 3d presents the measured (colormap) and theoretical (white dotted line) static dispersion relation of the CROW-PTC in the absence of dynamic modulation. In the theoretical model, we first consider the limit of an infinitely large DC-blocking capacitor. In this case, the eigenfrequency at $k = 0$ is nonzero and given by $\omega_{col} = 1/\sqrt{L_0 C_0}$, which corresponds to a global resonant mode. While at $k = \pi/a$, the eigenfrequency is given by the resonance frequency $\omega_{r0} = \sqrt{\omega_{col}^2 + 4/[(L_s - 2S)C_0]}$. When a large DC-blocking capacitor (up to 200 pF) is taken into account, only a slight correction to the dispersion curve is required (see the detailed theoretical derivation in Methods).

Periodic temporal modulation of the varactor capacitance at frequency $\Omega$ causes the *LC* resonance frequency to vary in time, leading to coupling between the static Floquet bands and opening a *k*-gap. Compared to the SSPP-PTCs, the CROW-PTCs benefit from the non-zero collective resonance frequency and the flatter static dispersion, which allows a larger expanded *k*-gap—even up to an infinite extent—to be realized.

When the resonance frequencies at the Brillouin zone center $k = 0$ ($\omega_{col}$) and the zone boundary $k = \pi/a$ ($\omega_{r0}$) are sufficiently close—specifically, the periodic temporal modulation satisfying $\omega_{col} < \Omega/2 < \omega_{r0}$—resonance effects are simultaneously excited at both wave vectors. This dual resonance maximizes the relative modulation depth of the effective permittivity, thereby generating a full *k*-gap that spans the entire momentum space. It is notable that the CROW-PTCs can always achieve a full $k$-gap provided the dynamic modulation depth is sufficiently large; the static bandwidth $\Delta\omega = \omega_{r0} - \omega_{col}$ dictates only the minimum modulation depth required for its formation. We define a resonator quality factor as $Q = \omega_{col} C_0 Z_0$, with $Z_0$ the free-space impedance; and an inter-resonator coupling strength $\kappa = 1/L_s$, with $L_s$ the effective inductance of the grooves on the top PCB layer (see Methods for the detailed definition of $Q$ and $\kappa$). Figure 3e presents a phase diagram illustrating the dependence of the $k$-gap size on $Q$ and $\kappa$ by changing $C_0/L_0$ and $L_s$ while fixing $\omega_{col} = 315$ MHz and modulation depth $\Delta C/C_0 = 0.18$. We observe that increasing $Q$ and reducing $\kappa$ narrows the static CROW-PTC bandwidth, which in turn enlarges the $k$-gap and facilitates the attainment of a full $k$-gap. Given practical constraints on modulation depth, we deliberately engineered a high-$Q$, weakly coupled CROW-PTC with a small static bandwidth $\Delta\omega$, enabling the realization of a full $k$-gap even under weak temporal modulation.

In the experiment, the modulation frequency was fixed at $\Omega = 700$ MHz, and the capacitance modulation depth reached $\Delta C = 5$ pF. Excitation of the CROW-PTC was achieved by injecting a Gaussian wave packet at the central position, and the corresponding band structure was obtained through spatiotemporal Fourier transformation, as presented in Fig. 3f. The measured full $k$-gap (colormap) manifests as an almost perfectly flat band in momentum space at the temporal Brillouin zone boundary $\omega = \Omega/2$, in excellent agreement with the theoretical dispersion (white dotted line). This flat band indicates that the associated eigenmodes are strongly localized in real space at arbitrary positions, with the field predominantly confined to the unit cell containing the excitation source. Concurrently, the nonzero imaginary part of the eigenfrequency induced by the dynamic modulation yields exponential temporal amplification of this tightly localized mode, as shown in Figs. 3g-i. By repositioning the excitation source (yellow stars) to different spatial positions, we observe that the spatiotemporal modes of the CROW-PTC remain tightly bound to the source location and undergo continuous amplification following the onset of dynamic modulation (white dashed lines), thereby constituting a non-propagating gain mode. Compared with the spatiotemporal field distributions of the SSPP-PTC displayed in Figs. 2h-j, the CROW-PTC spatiotemporal fields grow more rapidly and exhibit enhanced spatial localization, a direct consequence of the larger full $k$-gap. The coexistence of extremely spatial localization and sustained temporal amplification establishes the CROW-PTC as a promising platform for achieving lasing at arbitrary positions.

**Robustness of the expanded and full $k$-gap**

Finally, we explore the robustness of the expanded and full $k$-gaps against temporal disorder. Given that the relative capacitance modulation depths required to generate the two bandgap types differ experimentally, we define a disorder parameter that characterizes fluctuations in the instantaneous modulation depth relative to its mean. In the disorder-free case, the modulation depth is constant, $\Delta C/C_0 = \delta$. Under temporal disorder of strength $d$, the instantaneous modulation depth at each temporal step is randomly drawn from the uniform interval $[(1-d)\delta, (1+d)\delta]$, while the average modulation depth remains $\delta$. The maximum disorder range is therefore $[0, 2\delta]$. This parameterization permits a direct comparison of $k$-gap robustness at a given relative disorder level $d$.

We investigate three distinct temporal modulation regimes with disorder strengths $d = 0.2, 0.6,$ and $1.0$, as illustrated in Figs. 4a-c. The measured spatiotemporal distributions of $|H_z|^2$ for the expanded $k$-gaps in SSPP-PTCs and the full $k$-gaps in CROW-PTCs, under varying degrees of temporal disorder, are presented in Figs. 4d-f and Figs. 4g-i, respectively. With increasing disorder strength, the expanded $k$-gaps in the SSPP-PTCs (Figs. 4d-f) are systematically eroded by temporal disorder. At the highest disorder level ($d = 1$), temporal amplification within the expanded $k$-gap becomes virtually imperceptible, and spatially localized modes are no longer observed (Fig. 4f). In striking contrast, the full $k$-gaps in the CROW-PTCs (Figs. 4g-i) remain largely unperturbed; spatial localization and temporal amplification persist robustly even under strong disorder ($d = 1$). These results provide conclusive evidence that the full $k$-gap is substantially more robust than the expanded $k$-gap, implying a direct correlation between $k$-gap size and resilience to temporal disorder.

**Conclusion**

In summary, we have presented a systematic theoretical analysis of the formation mechanisms underlying three distinct types of $k$-gaps—normal, expanded, and full—revealing that their origins lie in the different responses of materials to external electric fields. These behaviors manifest as distinct Floquet modulations of the time-dependent permittivity. Both the expanded and full $k$-gaps arise from a resonance-enhanced mechanism: resonant responses effectively convert a finite parameter modulation into a giant effective modulation of the permittivity, thereby dramatically broadening the *k*-gap and, in the extreme case, extending it across the entire momentum space. Experimentally, we observe the first resonance-expanded *k*-gap in a SSPP-PTC with SSPP dispersion—and, more strikingly, the first full *k*-gap in a CROW-PTC governed by CROW dispersion. Direct spatiotemporal Fourier measurements resolve, for the first time, the complete structures of both the expanded and full *k*-gaps. We further demonstrate that a larger *k*-gap yields tighter spatiotemporal field confinement and greater robustness against temporal disorder. More intriguingly, we find that the CROW-PTCs, which exhibit full *k*-gaps, support arbitrary extreme spatial localization accompanied by temporal amplification, pointing to new opportunities for unconventional laser mechanisms rooted in PTCs.

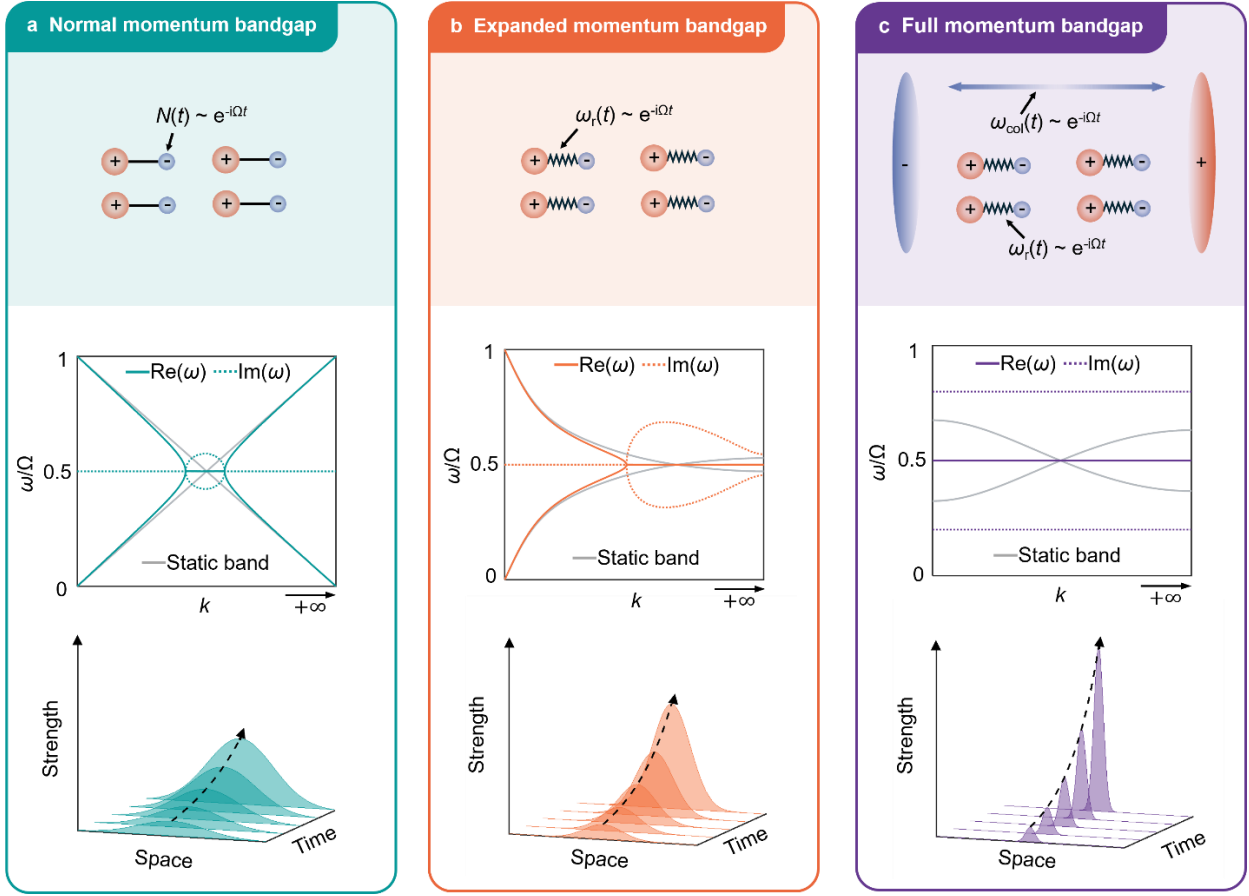

**Fig. 1 | Three different types of PTCs with normal, expanded, and full $k$-gaps. a**, Top panel: an ideal non-dispersive medium, in which electrons (grey sphere) and atomic nuclei (red sphere) are connected by rigid rods, resulting in an instantaneous response to an external electric field. In this case, periodic modulation of the electron density forms a PTC with a normal $k$-gap. Middle panel: band structure of the PTC with a normal (finite) $k$-gap. The cyan solid (dotted) line represents the real (imaginary) part of the band structure, and the grey solid line presents the dispersion of the ideal non-dispersive medium without time modulation. Bottom panel: the spatiotemporal field distributions within the normal $k$-gap. **b**, Top panel: a Lorentz-type dispersive medium, in which electrons and atomic nuclei are connected by elastic rods. When the frequency of the external electric field matches the resonance frequency of the Lorentz-type dispersive medium, resonant enhancement occurs. Middle panel: band structure of the PTC with an expanded (semi-infinite) $k$-gap. The orange solid (dotted) line represents the real (imaginary) part of the band structure, and the grey solid line presents the dispersion of the Lorentz-type dispersive medium without time modulation (SSPP dispersion). Bottom panel: the spatiotemporal field distributions within the expanded $k$-gap. **c**, Top panel: a Drude–Lorentz-type dispersive medium, in which electrons and atomic nuclei are connected by elastic rods, meanwhile the electrons are also driven by a background potential, giving rise to global longitudinal charge oscillations. Middle panel: band structure of the PTC with a full (infinite) $k$-gap. The purple solid (dotted) line represents the real (imaginary) part of the band structure, and the grey solid line presents the dispersion of the Drude–Lorentz-type dispersive medium without time modulation (CROW dispersion). Bottom panel: the spatiotemporal field distributions within the full $k$-gap.

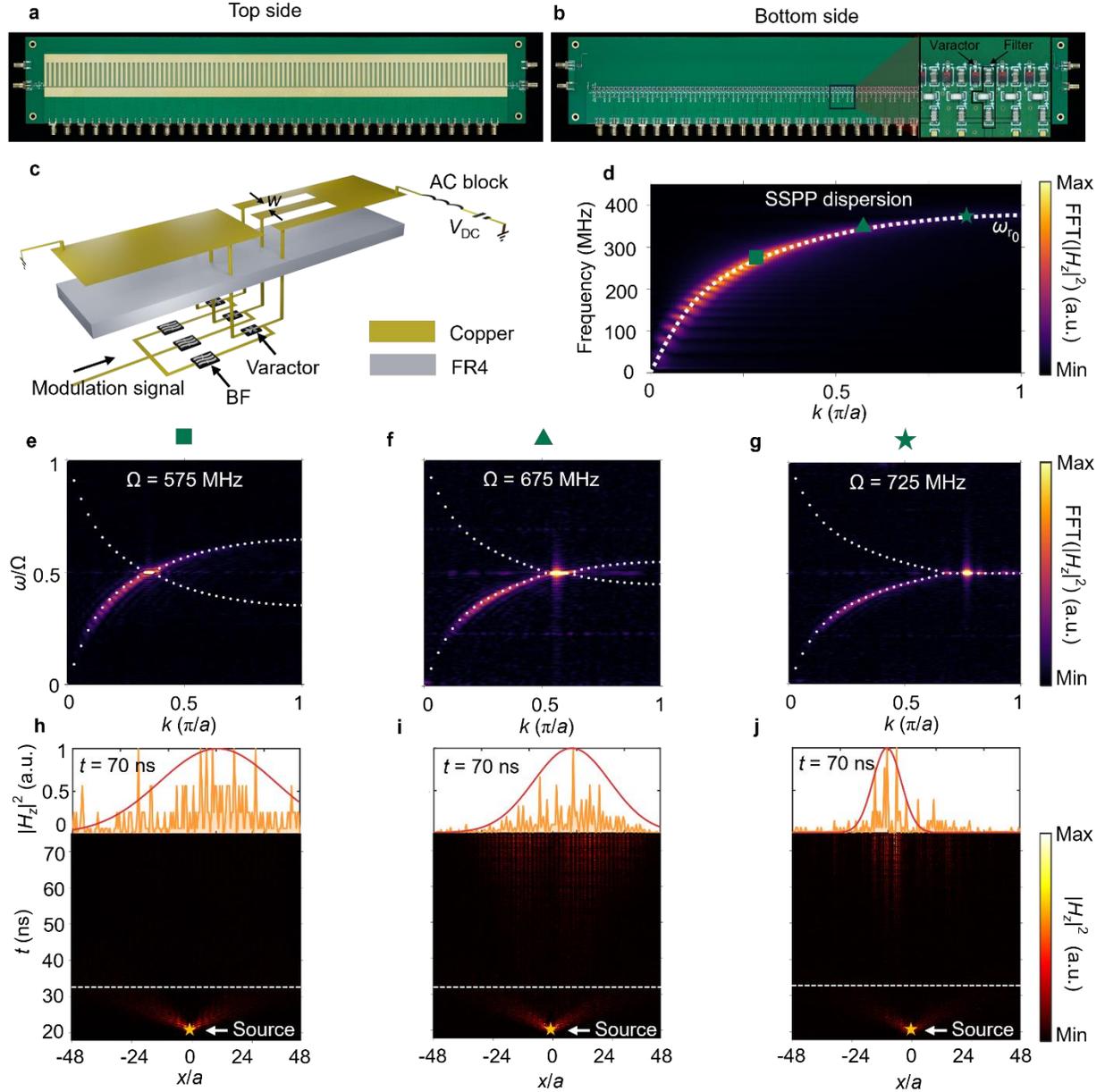

**Fig. 2 | Observation of expanded *k*-gap in a SSPP-PTC. a**, **b**, Photographs of the fabricated SSPP-PTC. The inset shows a magnified view of the varactor diode and filtering network. **c,** Schematic diagram of the unit cell of the SSPP-PTC. **d**, Measured (colormap) and theoretical (white dotted line) static dispersion relation of the microwave SSPP transmission line, which terminates at the spatial Brillouin zone boundary and exhibits a resonance frequency $\omega_{r0}$. **e–g,** Measured (colormap) and simulated (white dotted line) expanded *k*-gaps through resonance with different modulation frequencies $\Omega$. As $\Omega/2$ gradually approaches $\omega_{r0}$, the *k*-gap progressively broadens. **h–j**, Spatiotemporal distributions of $|H_z|^2$ in the three expanded *k*-gaps. A wider *k*-gap leads to modes that are more spatially localized and exhibit faster temporal amplification. The upper insets plot the measured (orange line) and fitted (red line) magnetic intensity distributions at *t* = 70 ns. The white dashed lines indicate the beginning of the dynamic modulation.

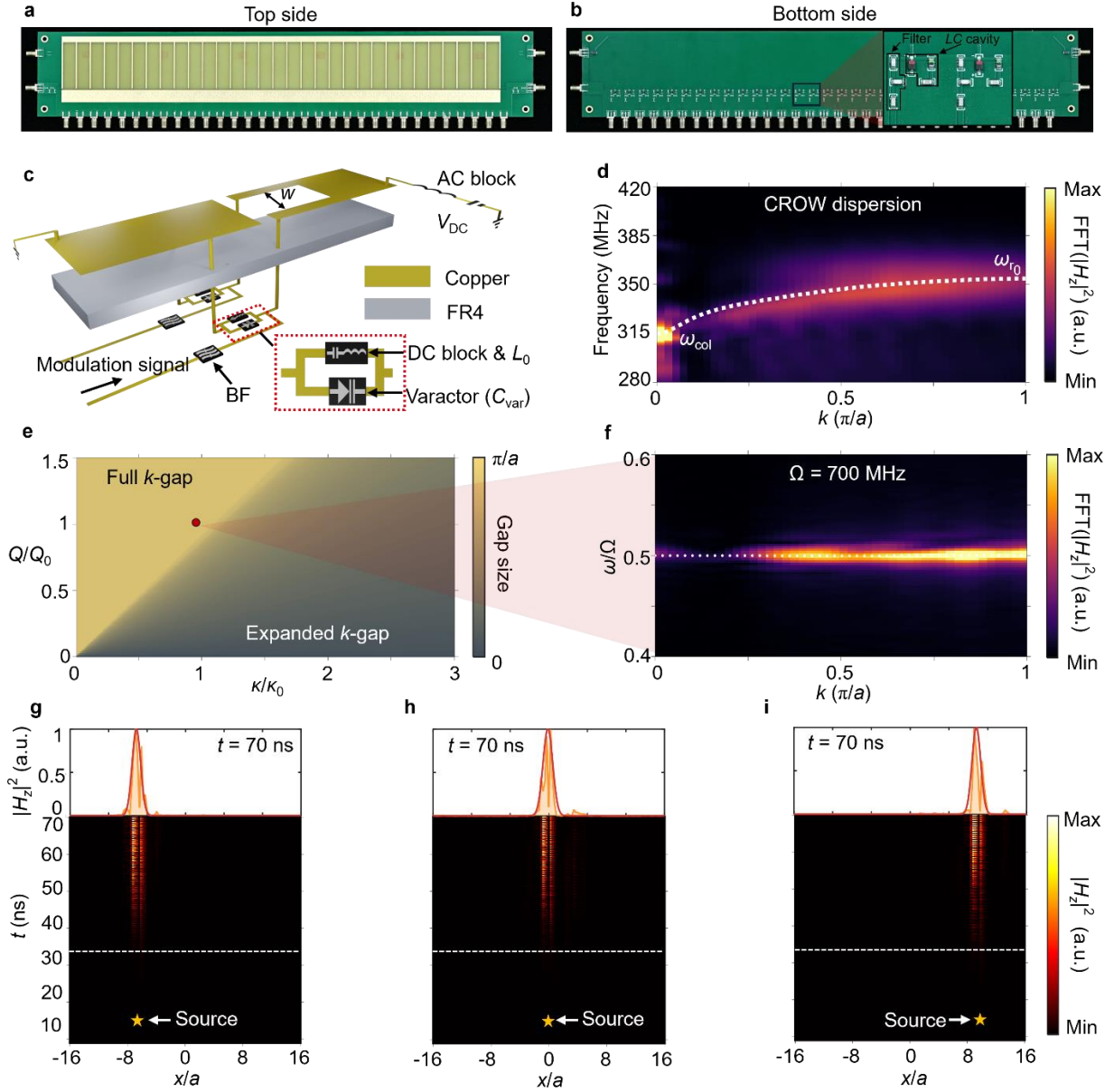

**Fig. 3 | Observation of full *k*-gap in a CROW-PTC. a**, **b**, Photographs of the fabricated CROW-PTC. The inset shows an enlarged view of the filter and LC cavity. **c**, Schematic diagram of a unit cell of the CROW-PTC. The inset shows a parallel-connected $L_0$ and $C_{\text{var}}$. **d**, Measured (colormap) and theoretical (white dotted line) static dispersion relation of the CROW-PTC, exhibiting two resonance frequencies $\omega_{\text{col}}$ at $k = 0$ and $\omega_{r0}$ at $k = \pi/a$, respectively. **e**, The *k*-gap size as a function of the resonator quality factor $Q$ and the inter-resonator coupling strength $\kappa$, where $Q_0 = 20.7$ and $\kappa_0 = 1/165 \text{ nH}^{-1}$. **f**, Measured (colormap) and theoretical (white dotted line) band structure of the CROW-PTC, exhibiting a full *k*-gap characterized by a flat band across the entire Brillouin zone at $\omega = \Omega/2$. **g–i,** Spatiotemporal distributions of $|H_z|^2$ in the full *k*-gap, with excitation sources (yellow stars) placed at different spatial positions, show that the CROW-PTC tightly localizes the mode at the excitation site and induces exponential temporal amplification. The upper insets plot the measured (orange line) and fitted (red line) magnetic intensity distributions at $t = 70$ ns. The white dashed lines indicate the onset of the dynamic modulation.

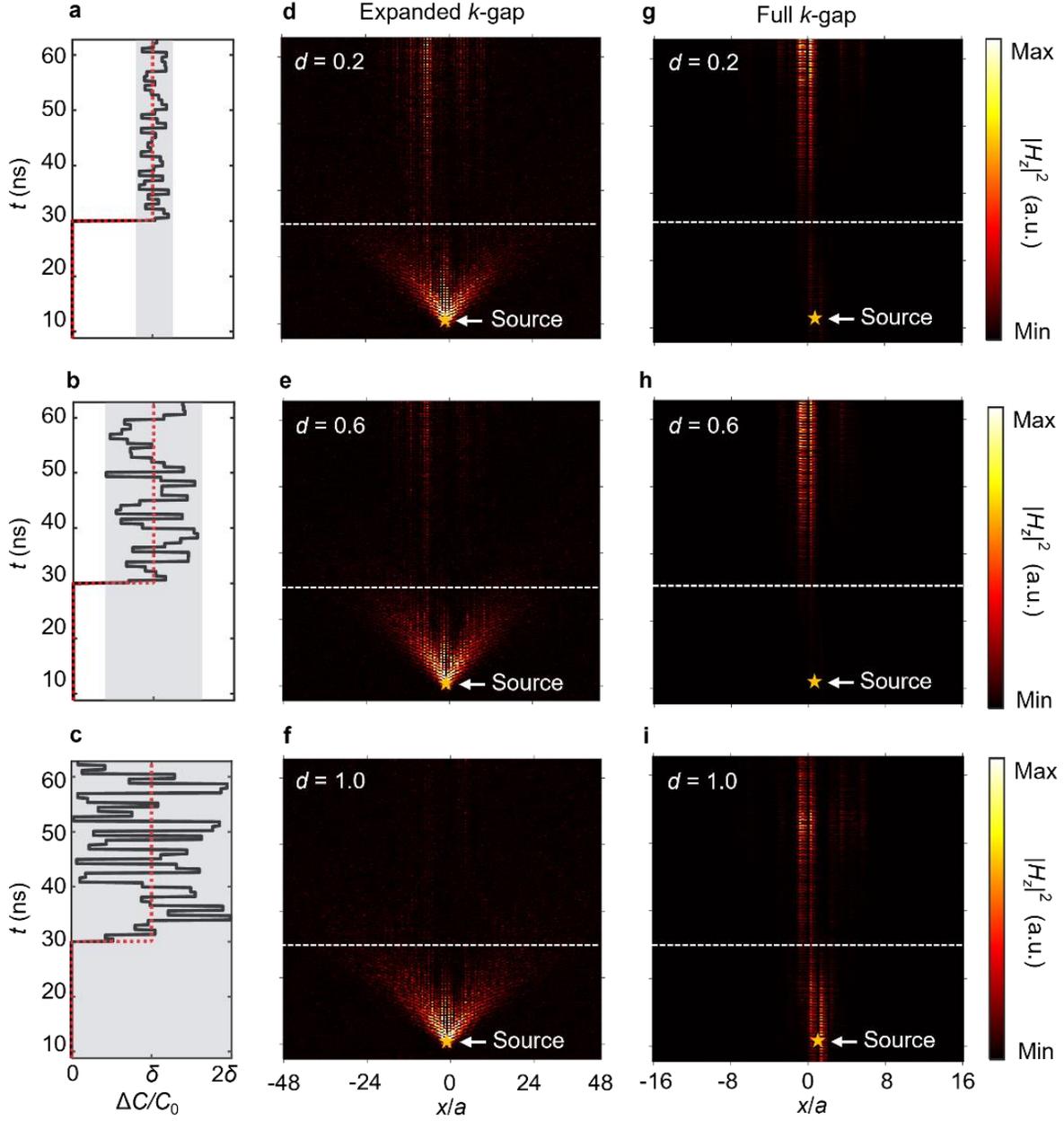

**Fig. 4 | Robustness of the expanded and full *k*-gaps. a-c**, Temporal disorders with different strengths, where the relative modulation depth randomly fluctuates within the interval $[(1-d)\delta, (1+d)\delta]$ while maintaining an average value of $\delta$. The disorder strengths are set to $d = 0.2$ in **a**, 0.6 in **b**, and 1 in **c**, respectively. The grey shaded region represents the disorder range. **d-f**, Spatiotemporal distributions of $|H_z|^2$ in the SSPP-PTC with different temporal disorder strengths, indicating that strong temporal disorder destroys the expanded *k*-gap, and the field amplification in the PTC is nearly unobservable. The white dashed lines indicate the beginning of the dynamic modulation. **g-i**, Spatiotemporal distributions of $|H_z|^2$ in the CROW-PTC with different temporal disorder strengths, showing that the full *k*-gap remains intact and exhibits strong robustness against the temporal disorder.

**Data availability**

The data that support the findings of this study are available from the corresponding author upon reasonable request.

**Acknowledgments**

Z.G. acknowledges funding from the National Key R&D Program of China (grant No. 2025YFA1412300), National Natural Science Foundation of China (grant No. 62361166627 and 62375118), Guangdong Basic and Applied Basic Research Foundation (grant No. 2024A1515012770), Shenzhen Science and Technology Innovation Commission (grant No. 20230802205352003), and High-level Special Funds (grant No. G03034K004). Z.Z. acknowledges the funding from the China Postdoctoral Science Foundation (grant No. 2025M773439).

**Authors Contributions**

Z.G. initiated the project; B.H. developed the theory and performed numerical calculations, with contribution from Z.Z.; Z.Z. designed and built the experimental setup; B.H. performed the experimental measurement, with contribution from G.Y.; B.H. and Z.Z. analyzed data; B.H., Z.Z., and Z.G. wrote and revised the manuscript; Z.G. supervised the project.

**Competing Interests**